\documentclass[journal]{IEEEtran}
\UseRawInputEncoding
\usepackage{cite}
\usepackage{amsmath,amssymb,amsfonts}
\usepackage{graphicx}
\usepackage{textcomp}
\usepackage{xcolor}
\usepackage{soul}
\usepackage{subcaption}
\usepackage{multirow}
\usepackage{siunitx}
\usepackage{makecell}
\usepackage{booktabs} 
\usepackage[usestackEOL]{stackengine}
\usepackage{amsmath}
\usepackage[linesnumbered,ruled]{algorithm2e}
\usepackage{algpseudocode}
\usepackage[euler]{textgreek}
\usepackage{svg}
\usepackage{hyperref}
\usepackage[symbol]{footmisc}
\usepackage{comment}
\usepackage{url}
\PassOptionsToPackage{hyphens}{hyperref}
\usepackage[hyphens]{hyperref}

\def\BibTeX{{\rm B\kern-.05em{\sc i\kern-.025em b}\kern-.08em
  T\kern-.1667em\lower.7ex\hbox{E}\kern-.125emX}}
\usepackage{xcolor}
\definecolor{darkgreen}{rgb}{0.0, 0.5, 0.0}
  
\newcommand{\blue}[1]{{{\color{blue} #1}}}

\newcommand{\violet}[1]{{{\color{violet} #1}}}

\renewcommand{\blue}{} 

\begin{document}
\pagenumbering{arabic}                
\pagestyle{plain}

\title{Low-Cost Sensing and Classification for Early Stress and Disease Detection in Avocado Plants}



\author{
    Abdulrahman~Bukhari,~\IEEEmembership{Student Member,~IEEE,}
    Bullo~Mamo, Mst Shamima~Hossain,
    Ziliang~Zhang,~\IEEEmembership{Student Member,~IEEE,}
    Mohsen~Karimi,
    Daniel~Enright,~\IEEEmembership{Student Member,~IEEE,}
    Patricia~Manosalva,
    Hyoseung~Kim,~\IEEEmembership{Member,~IEEE}
    \thanks{This work was supported by the USDA/NIFA grant (2020-51181-32198) and the UCR Delfino Agriculture Innovation Seed Fund.}    
    \thanks{A. Bukhari, Z. Zhang, M. Karimi, D. Enright, and H. Kim are with the Department of Electrical and Computer Engineering, University of California, Riverside, Riverside, CA 92521, USA (emails: \{abukh001, zzhan357, mkari007, denri006, hyoseung\}@ucr.edu).}
    \thanks{MS. Hossain is with the Department of Computer Science, University of California, Riverside, Riverside, CA, 92521, USA (email: mhoss037@ucr.edu).}
    \thanks{B. Mamo and P. Manosalva are with the Department of Microbiology and Plant Pathology, University of California, Riverside, Riverside, CA 92521, USA (emails: \{bullo.mamo, patricia.manosalva\}@ucr.edu).}   
}

\IEEEtitleabstractindextext{%

\begin{abstract}
With rising demands for efficient disease and salinity management in agriculture, early detection of plant stressors is crucial, particularly for high-value crops like avocados. This paper presents a comprehensive evaluation of low-cost sensors deployed in the field for early stress and disease detection in avocado plants. Our monitoring system was deployed across 72 plants divided into four treatment categories within a greenhouse environment, with data collected over six months. While leaf temperature and conductivity measurements, widely used metrics for controlled settings, were found unreliable in field conditions due to environmental interference and positioning challenges, leaf spectral measurements produced statistically significant results when combined with our machine learning approach. For soil data analysis, we developed a two-level hierarchical classifier that leverages domain knowledge about treatment characteristics, achieving 75-86\% accuracy across different avocado genotypes and outperforming conventional machine learning approaches by over 20\%. In addition, performance evaluation on an embedded edge device demonstrated the viability of our approach for resource-constrained environments, with reasonable computational efficiency while maintaining high classification accuracy. Our work bridges the gap between theoretical potential and practical application of low-cost sensors in agriculture and offers insights for developing affordable, scalable monitoring systems.

\end{abstract}

\begin{IEEEkeywords}
IoT, embedded edge devices, time-series classification, plant diseases, stress detection
\end{IEEEkeywords}

}



\maketitle
\IEEEdisplaynontitleabstractindextext
\IEEEpeerreviewmaketitle


\section{Introduction}
Plant diseases and environmental stressors pose significant challenges to agricultural productivity, particularly for high-value crops like avocados. Early detection of these issues is crucial for preventing widespread damage, e.g., leaf tip burns, and optimizing management strategies. Traditional methods for assessing plant health typically involve time-consuming in-lab analysis of soil and leaf samples, which require expensive benchtop equipment and specialist expertise. While high-end sensor technologies have shown promise in agriculture, their prohibitive costs limit widespread adoption, especially in resource-limited settings.



Driven by advances in wireless sensor networks, low-cost commercial sensors have emerged as a promising solution for agricultural monitoring due to their affordability and ease of integration with embedded systems. These sensors
can measure various parameters like soil moisture, electrical conductivity (EC), and leaf reflectance at a fraction of the cost of traditional equipment. However, despite their potential, there remains a critical gap in understanding their real-world performance and limitations when deployed to the field, particularly for early stress detection.


Our field study reveals this knowledge gap, particularly when applying established agricultural analysis techniques to low-cost sensors. Traditional analysis techniques, while effective with high-end laboratory equipment, often fail to produce statistically significant results when applied to data from low-cost sensors in the field due to inherent limitations in environmental interference and resulting sensing accuracy. For example, commonly used spectral indices for salinity estimation, such as the Normalized Difference Salinity Index (NDSI)~\cite{KHAN200596}, showed poor performance when applied to data from our low-cost spectrometer. 
This highlights the need for new analytical approaches specifically designed to address the unique challenges posed by low-cost sensors.


To address these challenges, we present a comprehensive study of low-cost sensors for detecting early signs of stress and disease in avocado plants, specifically focusing on salinity stress and Phytophthora root rot (PRR), which are major threats to avocado production~\cite{belisle2019new,celis2018salt}. We identify the limitations of these sensors and propose a novel approach to detecting stress and disease by strategically applying and integrating existing analytical and classification methods. Our contributions lie in system-level design, practical deployment in real-world conditions, and developing techniques to overcome data quality issues inherent in low-cost sensing in the field. This work bridges the gap between the theoretical potential of low-cost sensors and their practical application in agriculture, offering insights into the development of affordable, scalable monitoring systems for resource-limited settings. This paper makes the following key contributions:
\begin{itemize}
\item \textbf{System design:} We implemented a low-cost monitoring system using leaf and soil sensors across 72 avocado plants organized into four treatment groups. Over six months, this setup collected more than 800,000 sensor measurements, replacing the need for traditional in-lab analysis and enabling in-field detection of stress and disease in avocado plants (Section \ref{sec:system_design}).


\item \textbf{Stress and disease detection:} We focused on early signs of stress and disease detection by analyzing data recorded from two different locations: leaf and soil. 

(i) \textbf{Leaf spectral reflectance analysis.} We developed a novel approach for stress and disease detection that overcomes the limitations of low-cost spectrometers by applying multivariate pattern analysis with permutation testing. This approach achieves statistically significant results (with accuracy up to 89\%) in detecting stressed plants, while traditional spectral analysis methods~\cite{KHAN200596, abbas,DAUGHTRY2000229} fail to draw meaningful distinctions. (Section \ref{sec:leaf_analysis}).

(ii) \textbf{Soil EC and moisture analysis.} We proposed a two-level hierarchical classification approach tailored for early identification of disease and stress in different avocado genotypes. This approach achieves 75-86\% accuracy, substantially outperforming existing ML and time-series methods~\cite{Wu,Cliff,Xing,karim2019multivariate,wang2017time} by over 20\% (Section \ref{sec:soil_analysis}).



\item \textbf{Observations:} We discovered that leaf temperature and leaf EC measurements are unreliable when data come from low-cost sensors deployed in the field, despite their proven effectiveness with high-end sensors in the literature. We present our discussion in Section \ref{sec:discussion}.
\end{itemize}

\section{Related Work}
\subsection{Agricultural Monitoring Technologies}

Various plant monitoring systems have been studied in agricultural literature to assess plants' health~\cite{Chabrillat, Zhang, Wang2023, Sikka, vasisht2017farmbeats}. For large-scale environments where no direct contact with plants is available, drone-based imaging~\cite{Boubin,mavic3_agri,dslrpros_drones} and satellite imaging~\cite{Allbed,agronomy13030716} have been used to analyze spectral signatures to characterize soil composition and moisture levels. While effective in assessing overall health patterns across regions, these approaches are not only costly but also have limitations in identifying individual plants' health issues.




Sensors placed in direct contact with plant tissue or soil provide detailed plant-level data complementing the aforementioned satellite and drone imaging approaches.
Leaf sensors~\cite{Afzal001,Afzal002} measure the leaf's electrical conductivity or capacitance to estimate leaf thickness and moisture, which are strong indicators of plant health. However, these studies have been primarily conducted in controlled environments. In addition to electrical measurements, spectrometers have been utilized to analyze leaf light reflectance~\cite{SILVA2023108001,ECARNOT201344}. However, high-end spectrometers used in previous work, such as Agilent 240FS AA~\cite{agilent_240fs}, are costly and often require a controlled environment (e.g., in-lab settings) to prevent external light interference, limiting their applicability for practical use.



Soil sensors also play a vital role by providing data on soil conditions such as moisture level, salinity, and nutrient levels~\cite{Luong,chen}. 
High-end soil sensors, including those from companies like Decagon and Sentek~\cite{decagon_site,sentek_site,tdr_fdr_review}, employ time-domain reflectometry (TDR) and frequency-domain reflectometry (FDR) for precise measurements of soil moisture content. However, they are relatively expensive compared to their low-cost alternatives.

In summary, these high-accuracy sensors are predominantly used in controlled research environments with a limited number of plants, where their cost and labor demands are more manageable. Although affordable, low-cost sensors are becoming increasingly available, there is limited research evaluating their reliability in real-world agricultural settings or optimizing analytical methods to address their inherent limitations.

\subsection{Sensors Data Analysis and Classification}\label{sec:related_work_analysis}
Various approaches have been explored in agricultural research to analyze leaf sensor data for detecting plant stressors, such as salinity. Spectrometer analysis based on leaf light reflectance has been widely adopted for estimating salinity stress through spectral indices like the Normalized Difference Salinity Index (NDSI), Normalized Index (NI), and Salinity Index 1 (N1) \cite{KHAN200596, abbas, DAUGHTRY2000229}. These indices capture changes in light reflectance at specific wavelengths, which correlate with salinity stress responses. 

Traditional machine learning techniques, including Random Forests, K-Nearest Neighbors (KNN), and Support Vector Machines (SVM)~\cite{Wu,Cliff,Xing,Hunt1989DetectionOC}, have been effectively applied to classify time-series data,  while deep learning methods, such as recurrent neural networks (RNN) and residual neural networks (ResNet)~\cite{karim2019multivariate,wang2017time}, have demonstrated greater potential by capturing temporal dependencies in sensor data. Emerging techniques like Multivariate Unsupervised Symbols and dErivatives (MUSE)~\cite{schafer2017multivariate} have also been explored to further improve classification accuracy. 
However, these methods can be challenged when overlapping classes are present in the dataset, which can significantly degrade their performance, as will be shown in Sec.~\ref{sec:soil_analysis}.



Several state-of-the-art methods have been developed for the classification of time-series data. The MR-HYDRA method combines key aspects of convolutional kernels and dictionary-based techniques, offering faster and more accurate results than traditional approaches~\cite{hydra}. The HIVE-COTE 2.0 method incorporates novel classifiers like Temporal Dictionary Ensemble (TDE) and Diverse Representation Canonical Interval Forest (DrCIF), making it highly effective for real-time, complex sensor data classification in agricultural systems~\cite{hive}. Additionally, the QUANT method simplifies interval-based classification by using quantiles and fixed intervals, achieving high accuracy with minimal computational requirements, making it a highly efficient approach for large-scale time series tasks~\cite{quant}. These methods represent the cutting edge in time series classification, applicable to soil sensor data.

\begin{table}[!th]
\centering
\caption{Cost breakdown of agricultural sensing systems}
\label{tab:cost}
\resizebox{\linewidth}{!}{%
\begin{tabular}{|c|l|c|}
\hline
\textbf{System Type}                                               & \multicolumn{1}{c|}{\textbf{Equipment}}                                                                                                                                         & \textbf{\begin{tabular}[c]{@{}c@{}}Initial Cost (USD)\end{tabular}}                                                   \\ \hline
\begin{tabular}[c]{@{}c@{}}Drone-based \\Imaging \end{tabular}     & \begin{tabular}[c]{@{}l@{}}DJI Mavic 3 Multispectral\\ or Matrice 350 RTK\end{tabular}                                                                                          & \$9,900--\$35,000  \\ \hline

\begin{tabular}[c]{@{}c@{}}Leaf Spectral \\ Analysis\end{tabular}   & Agilent 240FS AA, lab-grade AAS                                                                                   & \$20,000  \\ \hline
\begin{tabular}[c]{@{}c@{}}Soil Analysis\end{tabular} & \begin{tabular}[c]{@{}l@{}}High-end TDR/FDR-based soil \\ moisture sensors\end{tabular} & \begin{tabular}[c]{@{}l@{}}\$200--\$1,000 per plant \end{tabular} \\ \hline
\multirow{3}{*}{\begin{tabular}[c]{@{}c@{}}Proposed \\ Solution\end{tabular}}   & \begin{tabular}[c]{@{}l@{}}Low-cost soil sensors \\ (EC, moisture) with Bluetooth\end{tabular}                                                                                  & \$25 per plant   \\ \cline{2-3}
& \begin{tabular}[c]{@{}l@{}}Handheld leaf sensor with \\ spectrometer, IR array, and EC pad\end{tabular}                                                                                  & \$200--\$320 \\ \hline
\end{tabular}
}
\end{table}

\begin{figure*}[!th]
  \centering
  \includegraphics[width=1\linewidth]{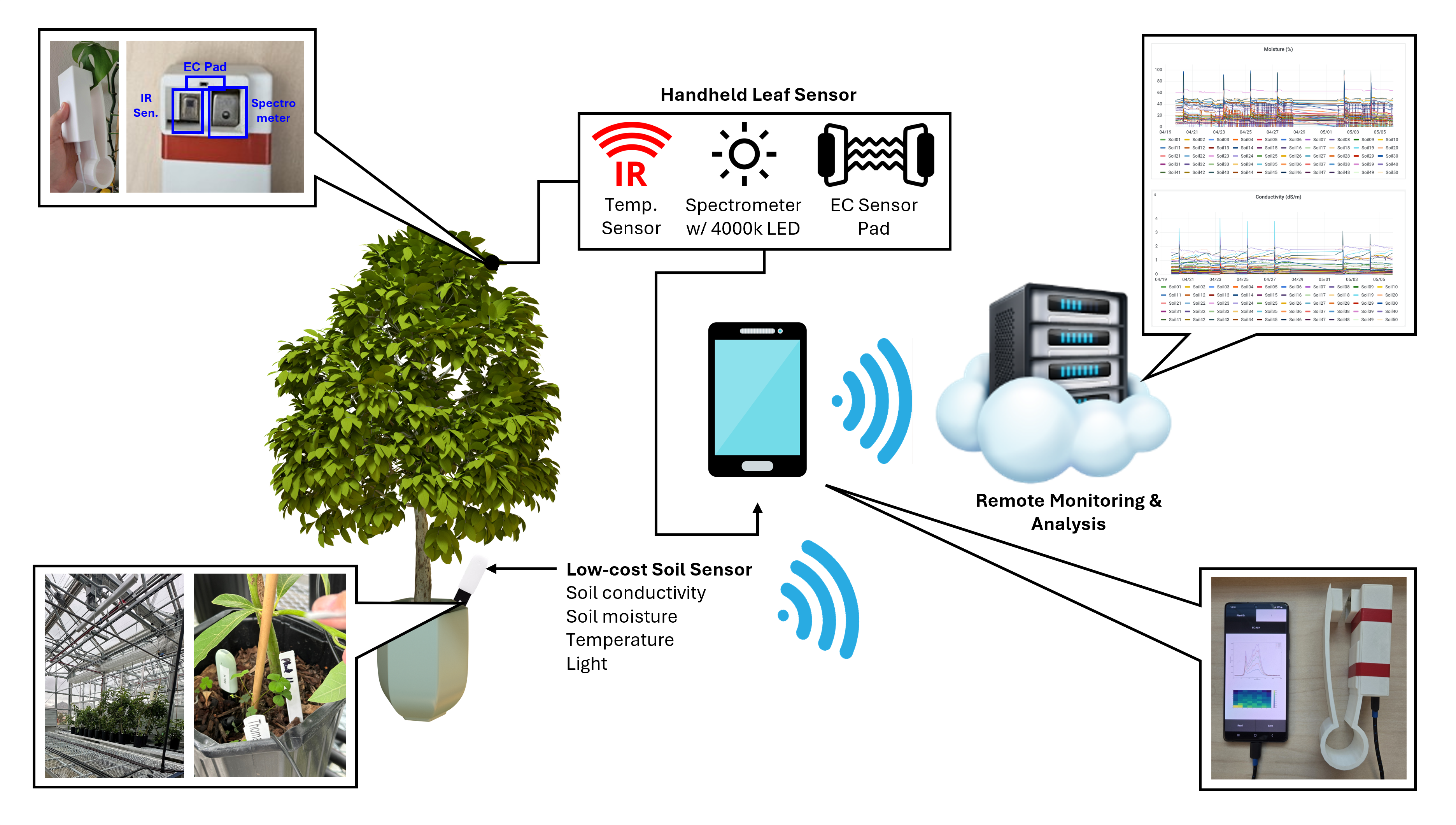}
  \caption{Greenhouse Plant Monitoring System}
  \label{fig:framework}
\end{figure*}

\section{Monitoring System Design} \label{sec:system_design}
This section describes our plant monitoring system. We deployed our system in a greenhouse at UC Riverside, CA, USA, where we conducted a comprehensive six-month study of 72 avocado plants to validate its effectiveness.

\subsection{Contrast to Existing Literature}

Table~\ref{tab:cost} provides a detailed breakdown of costs associated with various agricultural sensing systems, including our proposed low-cost solution.
Low-cost soil and leaf sensors offer a promising, affordable alternative, particularly for stress detection through monitoring key factors like conductivity and moisture at the root zone to detect issues such as salinity and Phytophthora root rot (PRR). However, existing studies largely focus on high-end systems, leaving a gap in understanding the performance and limitations of low-cost solutions in real-world agricultural settings. Our work addresses this gap by investigating whether statistically significant plant health assessments can be achieved combining these cost-effective solutions with existing data analysis and classification techniques.

\section{Monitoring System Design} \label{sec:system_design}
This section describes our plant monitoring system. We deployed our system in a greenhouse at UC Riverside, CA, USA, where we conducted a comprehensive six-month study of 72 avocado plants to validate its effectiveness.

\subsection{System Overview}

Our system aims to enable early detection of salinity stress and PRR disease in avocado plants through low-cost sensor measurements. 
The system architecture, shown in Fig.~\ref{fig:framework}, consists of three main components.



\begin{itemize} 
\item Leaf Monitoring: Uses a portable leaf sensor device to measure spectral light reflectance, temperature, and leaf conductivity. 
\item Soil Monitoring: Equips each avocado plant pot with a low-cost soil sensor that continuously measures electrical conductivity (EC) and moisture levels. 
\item Data Acquisition: Stores collected sensor data on a remote server for preprocessing and analysis. \end{itemize}





\subsection{Leaf Monitoring}

Continuous monitoring of multiple leaves from multiple plants is not practically doable because we cannot attach to each leaf all the sensors needed to measure spectral reflectance, temperature, and conductivity. Hence, we opted to design a custom handheld leaf sensor device that a user can carry to the greenhouse and take measurements of select samples.

Our handheld leaf sensor prototype, as shown in the left top corner of Fig.~\ref{fig:framework}, integrates three types of sensors in a 3D-printed housing to measure key physiological parameters of avocado leaves: (i) a micro-spectrometer for measuring leaf surface reflectance, (ii) an IR sensor for leaf temperature measurement, and (iii) an electrical conductivity (EC) pad designed to assess moisture content by pinching the leaf vein. To measure leaf reflectance, we used the Hamamatsu C12880MA MEMS micro-spectrometer, which
captures 288 data points across a spectral range of 340-850 nm with 15 nm resolution. 


For leaf temperature measurement, we utilized the Panasonic AMG8833 IR array sensor \cite{Panasonic2024}, which allows us to take an 8x8 array of thermal measurements across the leaf surface. 


Lastly, we custom-designed a small circuit as an EC sensor to measure the leaf's conductivity, with a measurement range of 10 to 400 nS/cm. This circuit includes an LM10 operational amplifier \cite{TexasInstruments2024} as a buffer to enhance signal stability. 

The circuit consists of a 10 $M\si{\ohm}$ resistor ($R1$) and a second variable resistor ($R2$) that changes with leaf conductivity. The Leaf EC is calculated as:
$EC\: in\: S/m = \frac{L}{R2\times A}$
where $L$ is the distance between electrodes (0.25 cm), and $A$ is the cross-sectional area (0.01 cm$^2$) of the electrode. 

\subsection{Soil Monitoring}

Unlike leaf measurements that require manual sampling, soil conditions can be continuously monitored by installing sensors directly to each plant's pot. We chose to use a low-cost commercial soil sensor, priced at approximately \$25 per unit (left bottom corner of Fig.~\ref{fig:framework}). Each sensor unit has two prongs that penetrate the soil to capture EC and moisture values, while the upper portion houses additional components for ambient measurements which we did not use in our work.


Moisture level is detected by a capacitive sensor strip. 
For EC measurements, the sensor is equipped with four stainless steel electrodes to ensure reliable and long-term use. 
To enable continuous monitoring, each soil sensor transmits data wirelessly to an embedded edge device (Raspberry Pi) located in the greenhouse via Bluetooth 4.1.
The sensors are powered by CR2032 button cell batteries with an average lifespan of three months, allowing low maintenance for our experiments.

\begin{table*}[!th]
\centering
\footnotesize
\blue{
\caption{\blue{Summary of Sensor Configuration and Data Acquisition}}
\label{tab:sensor_summary}
\begin{tabular}{|c|c|c|c|c|c|}
\hline
\textbf{Sensor} & \textbf{Device Type} & \textbf{Location} & \textbf{Sampling Frequency} & \textbf{Data Acquisition} & \textbf{Synchronization} \\ \hline
Spectral & Handheld leaf sensor & Marked leaves & Weekly (bi-weekly initially) & Manual readings & Same leaf per session \\ \hline
Temperature & Handheld leaf sensor & Marked leaves & Weekly (bi-weekly initially) & Manual readings & Same leaf per session \\ \hline
Conductivity & Handheld leaf sensor & Marked leaves & Weekly (bi-weekly initially) & Manual readings & Same leaf per session \\ \hline
Moisture & Soil sensor & Root zone & Every 30 min & Raspberry Pi to InfluxDB  & Shared timestamps \\ \hline
Conductivity & Soil sensor & Root zone & Every 30 min & Raspberry Pi to InfluxDB  & Shared timestamps \\ \hline
\end{tabular}
}
\end{table*}

\subsection{Data Acquisition and Management}\label{data_aques}

Our system collects data at different rates depending on the type of sensor device. For the handheld leaf sensor, measurements were taken manually on specific dates since physiological changes in leaves develop gradually over time. During each measurement session, data were collected via an Android device (right bottom corner of Fig.~\ref{fig:framework}), labeled with the corresponding plant's rootstock and treatment group, and uploaded to the remote server. 


In contrast, the soil sensors provide continuous monitoring. We configured them to take readings every 30 minutes to capture changes in soil condition over time, which is particularly important around watering schedules. Once the Raspberry Pi receives data, it labels each measurement with its timestamp and corresponding plant information before forwarding it to the server. Then the server maintains it in the time-series database, InfluxDB, with the plant's information. 

\section{Experimental Datasets}
\label{sec:data}

This section summarizes leaf and soil sensor datasets used in our experiments. We collected six months of sensor data, from November 2023 to April 2024, from our greenhouse setup at the University of California, Riverside, containing 72 avocado plants representing three distinct rootstocks: (i) Thomas, (ii) PP40, and (iii) PP45. The Thomas rootstock is susceptible to both salinity and PRR; PP40 is resistant to PRR and tolerant to salinity; and PP45 is resistant to PRR but susceptible to salinity. 
Each rootstock is further divided into four treatment categories: (i) control, (ii) salinity, (iii) PRR, and (iv) salinity and PRR, \blue{with each rootstock-treatment combination including the same number of plants ($n = 6$). This grouping enables equal representation across rootstock-treatment combinations, eliminating class imbalance during model training and evaluation. Note that our analysis in this paper primarily focuses on the Thomas and PP40 rootstocks, since PP45 did not have noticeable changes from treatments due to its resistance.}



\blue{Below we provide detailed explanations on how we set up and obtained each type of sensor data. A summary of all sensors, sampling frequencies, and acquisition procedures is provided in Table~\ref{tab:sensor_summary} to provide a clear overview of the experimental setup and data acquisition process.}

\subsection{\blue{Dataset Construction}}

\smallskip \noindent \textbf{Leaf Data.} Leaf sensor datasets are organized into the dates of data collection sessions.
\blue{At each session, measurements were collected from marked leaves on all plants to ensure consistency over time. All measurements were conducted at the same time of day under controlled lighting conditions inside the greenhouse. For conductivity measurements, the handheld device was pinched onto the marked leaf surface to ensure a consistent contact point across all samples. Spectral and temperature readings were taken from the same marked leaves at fixed positions. Each reading was repeated three times per leaf at different points on the same leaf to ensure consistency and to capture stress variations.}
Data collection was initially conducted bi-weekly for the first three months to allow sufficient time for treatment effects to manifest on the leaves. As visible symptoms appeared more rapidly later in the study, data collection frequency was increased to weekly until the experiment’s conclusion. 



\smallskip \noindent \textbf{Soil Data.} For soil measurements, we first sliced the entire six months of time-series data into daily windows of 48 readings each (one reading every 30 minutes, as explained in Section~\ref{data_aques}. 
Then, we divided them into three distinct periods: (i) late December 2023 to late February 2024, when no visible changes such as tip burns were present on the leaves; (ii) late February to mid-March 2024, when some plants began showing symptoms; and (iii) mid-March to late April 2024, when significant tip burns were observed. Note that initial data points from November 2023 to mid-December 2023 were not used since they were noisy due to greenhouse climate and watering setup adjustments. 

\blue{\smallskip \noindent \textbf{Data Preprocessing.} To improve the reliability of soil moisture and conductivity data, we first removed any values outside the nominal range to eliminate random spikes and obvious sensor malfunctions. Duplicated samples (e.g., due to wireless packet retransmissions) were also removed using annotated timestamps. Then, linear interpolation was used to fill missing values and ensure continuity in the time-series data. A moving average smoothing filter (window size = 20) was also used to reduce transient fluctuations while preserving key trends. We did not use more aggressive outlier removal beyond the simple range check, as short-term variability within the nominal range may reflect physiological responses or irrigation events rather than sensor noise.}

\blue{\smallskip \noindent \textbf{Training/Testing Datasets and Labeling.} Data from each experimental period were split chronologically by date to prevent temporal leakage, with training and testing datasets using data from separate, non-overlapping time periods. Specifically, training data were collected from late December 2023 to mid-March 2024, and testing data from mid-March to late April 2024. Treatment group labels were assigned solely based on the experimental design, according to each plant’s rootstock and treatment group, and were not influenced by sensor readings or observed symptoms. Hence, there were no ambiguous or borderline cases in labeling.}

\subsection{\blue{Relevance to Stress and Disease Monitoring}}
\blue{The selected data types (soil moisture, soil electrical conductivity, and leaf spectral measurements) are closely linked to physiological responses in plants under salinity stress and PRR infection. 

\begin{itemize}
\item \textbf{Soil electrical conductivity (EC)} is directly influenced by salinity levels due to increased ion concentration in the soil. Additionally, both salinity and PRR affect root function, potentially altering soil EC dynamics through changes in water uptake and nutrient transport.

\item \textbf{Soil moisture} trends provide insight into plant water uptake efficiency. Salinity and PRR can both reduce root function, which in turn affects how moisture is drawn from the soil over time.

\item \textbf{Leaf spectral reflectance} captures physiological changes in the plant, such as reduced chlorophyll content, altered water status, and changes in cellular structure, all of which influence reflectance properties, particularly in the visible and near-infrared ranges. These spectral changes are well-documented indicators of stress in avocado and other crops.
\end{itemize}
Together, these measurements provide a comprehensive view of the plant-soil system’s response to stress, justifying their selection for this study.}

\section{Leaf Sensor Data Analysis and Classification} \label{sec:leaf_analysis}


This section presents our analysis focusing specifically on leaf spectral reflectance data. While our handheld leaf sensing device has collected multiple leaf measurements (spectral reflectance, temperature, and EC), we focus on spectrometer data here as it provided the most reliable results (leaf temperature and EC measurements are discussed in Section~\ref{sec:discussion}). We first discuss the limitations of existing analysis methods, and then present our approach and experimental results.

\subsection{Existing Methods}

Leaf reflectance analysis in agricultural studies typically relies on two established approaches: wavelength-wise statistical analysis~\cite{wang2009distinguishing} and calculation of spectral indices~\cite{KHAN200596,DAUGHTRY2000229,abbas}. These methods have proven effective with high-end spectrometers in controlled laboratory settings, particularly for detecting plant stress conditions like salinity. Hence, we chose these two as our baselines.


The wavelength-wise statistical approach uses Analysis of Variance (ANOVA) testing to identify significant differences between control and salinity groups based on the reflectance value of a single wavelength (SW) at a time~\cite{wang2009distinguishing}. Specifically:

\begin{itemize}
    \item \textbf{Null Hypothesis (H\textsubscript{0}):} There is no statistically significant difference in reflectance between treatment groups at each wavelength ($p > 0.05$).
    \item \textbf{Alternative Hypothesis (H\textsubscript{1}):} There is a statistically significant difference in reflectance between treatment groups at one or more wavelengths ($p \leq 0.05$).
\end{itemize}
However, when applied to our low-cost spectrometer data, this approach failed to produce statistical significance ($p > 0.05$) between different treatment groups for all wavelength ranges (340-850 nm), even when visible symptoms like tip burn were present. 



For the calculation of spectral indices, several methods have been developed specifically for salinity detection. The most widely used are: Normalized Difference Salinity Index (NDSI)~\cite{KHAN200596},
Normalized Index (NI)~\cite{DAUGHTRY2000229}, and Salinity Index~1 (N1)~\cite{abbas}. These indices combine reflectance values from multiple wavelengths to estimate salinity levels. For example, NDSI uses the formula:
\begin{equation}
\text{NDSI} = \frac{R_{665} - R_{842}}{R_{665} + R_{842}}
\label{eq:ndsi}
\end{equation}
where \( R_w \) represents the reflectance measured at wavelength \( w \) (in nanometers), with 665 nm and 842 nm being specific wavelengths used in the computation.  

Similarly, NI and N1 use different wavelength combinations:  
\begin{equation}
\text{NI} = \frac{R_{705} - R_{750}}{R_{705} + R_{750}}
\label{eq:ni}
\end{equation}
\begin{equation}
\text{N1} = \frac{R_{600} - R_{800}}{R_{600} + R_{800}}
\label{eq:n1}
\end{equation}
where \( R_{705} \), \( R_{750} \), \( R_{600} \), and \( R_{800} \) correspond to the reflectance values at 705 nm, 750 nm, 600 nm, and 800 nm, respectively.  


\begin{table}[!th]
\small
\centering
\caption{P-values from ANOVA tests on single-wavelength (SW), NDSI, NI and N1 estimations}
\label{tab:pvalues}
\begin{tabular}{|c|c|c|c|c|}
\hline
Date      & SW    & NDSI  & NI    & N1    \\ \hline
28 Mar 2024  & 0.685 & 0.573 & 0.495 & 0.436 \\ \hline
3 Apr 2024   & 0.683 & 0.329 & 0.936 & 0.547 \\ \hline
12 Apr 2024  & 0.227 & 0.050 & 0.437 & 0.063 \\ \hline
18 Apr 2024  & 0.463 & 0.351 & 0.173 & 0.041 \\ \hline
\end{tabular}
\end{table}

These indices, however, also failed to provide statistical significance between treatment groups. Table~\ref{tab:pvalues} 
shows the p-values obtained from ANOVA tests based on these indices. None of the indices achieved the standard significance threshold (\( p < 0.05 \)), except in a few isolated cases when symptoms were already severe (12 April 2024 and afterward). 
These results indicate that traditional analysis methods, while effective with high-end equipment, are not suitable for data from low-cost spectrometers.

\subsection{Our Approach: Multi-wavelength SVM}
Given the limitations of conventional methods with low-cost spectrometer data, we adopted a multivariate pattern analysis (MVPA) approach with a permutation testing framework. Rather than analyzing individual wavelengths or predefined spectral indices, our approach aims to capture complex patterns across all wavelengths in distinguishing between control and treatment groups.

For classification, we utilized a support vector machine (SVM) with a linear kernel, selected for its simplicity and effectiveness in identifying linear relationships between sensor features and the four groups. 
To ensure a robust evaluation of the model's performance, we applied stratified k-fold cross-validation, which preserves the distribution of treatment groups across folds.

To validate the statistical significance of our results, we performed permutation testing, a non-parametric test to determine whether the classification performance was statistically meaningful or occurred by chance. We define the hypotheses as follows:

\begin{itemize}
\item \textbf{Null Hypothesis (H$_0$)}: There is no association between input features (spectrometer data) and treatment labels (control vs. salinity vs. PRR). In this case, the classification model does not perform better than random chance, implying $p > 0.05$.
\item \textbf{Alternative Hypothesis (H$_1$)}: A significant association exists between the input features and treatment labels, meaning the classification model outperforms random chance with statistical significance ($p < 0.05$).
\end{itemize}

For each test iteration, we randomly shuffled the treatment labels and retrained the SVM model to generate a null distribution of accuracies. The p-value was computed by comparing the observed classification accuracy against this null distribution.

\subsection{Experimental Results}



We have already shown in the previous subsection that existing methods fail to draw statistically meaningful results for our data. However, one may wonder about their performance when combined with SVM as our approach does. To ensure a fair comparison, we introduce three new baselines: NDSI-SVM, NI-SVM, and N1-SVM, which are SVM models trained using NDSI, NI, and N1 indices, respectively. The single-wavelength (SW) index is not included here because it failed to show any meaningful differentiation between treatment groups.

We present two sets of experimental results to demonstrate the effectiveness of our approach. First, we evaluated model performance on the control and salinity groups through cross-validation using data collected on specific dates. For each date, we conducted stratified 5-fold cross-validation and performed 1,000 permutation tests to assess statistical significance.

\begin{table*}[!th]
\blue{
\centering
\footnotesize
\caption{Comparison of our approach against conventional spectral indices with SVM classification}
\label{tab:SVMlog}
\begin{tabular}{|c|cc|cc|cc|cc|cc|}
\hline
\multirow{2}{*}{\textbf{Date}} & \multicolumn{2}{c|}{\textbf{NDSI-SVM}} & \multicolumn{2}{c|}{\textbf{NI-SVM}} & \multicolumn{2}{c|}{\textbf{N1-SVM}} & \multicolumn{2}{c|}{\textbf{Our Approach (w/o PCA)}} & \multicolumn{2}{c|}{\textbf{Our Approach (PCA = 15)}} \\ \cline{2-11}
 & Accuracy & P-Value & Accuracy & P-Value & Accuracy & P-Value & Accuracy & P-Value & Accuracy & P-Value \\ \hline
28 Mar 2024 & 52.78\% & 0.578 & 55.56\% & 0.488 & 55.56\% & 0.438 & \textbf{61.7\%} & \textbf{0.006} & 63.3\% & 0.000 \\ \hline
3 Apr 2024 & 50.00\% & 0.326 & 58.33\% & 0.543 & 50.00\% & 0.929 & \textbf{72.8\%} & \textbf{0.000} & 73.9\% & 0.000 \\ \hline
12 Apr 2024 & 72.22\% & 0.050 & 52.78\% & 0.444 & 61.11\% & 0.063 & \textbf{84.4\%} & \textbf{0.000} & 85.0\% & 0.000 \\ \hline
18 Apr 2024 & 47.22\% & 0.350 & 61.11\% & 0.168 & 52.78\% & 0.041 & \textbf{89.4\%} & \textbf{0.000} & 77.2\% & 0.000 \\ \hline
\end{tabular}
}
\end{table*}

As shown in Table~\ref{tab:SVMlog}, our approach achieved consistently high accuracy (up to 89\%) with strong statistical significance ($p < 0.05$) during the later stages of the experiment when stress symptoms became visible. Since leaf measurements correspond to discrete data points collected on specific dates, classification accuracy is reported for each individual date as shown in this table. 


As the results indicate, spectral indices alone did not provide strong discriminatory power under the tested conditions. Although some combinations, such as NDSI-SVM on April 12, showed moderate performance, overall accuracy and statistical significance were lower compared to our multi-wavelength approach.

\blue{To assess the potential impact of dimensionality on model performance, we conducted additional experiments using Principal Component Analysis (PCA) to reduce the 288 spectral channels to 15 principal components. The choice of 15 principal components was based on testing a range of 1 to 20 components, where 15 yielded the highest cross-validation accuracy on our training set. As shown in Table~\ref{tab:SVMlog}, PCA did not consistently improve classification accuracy across all test dates. These findings suggest that the original feature space of our linear SVM does not suffer from the curse of dimensionality or overfitting, and that PCA-based dimensionality reduction offers no clear advantage for this specific application.}

\begin{table}[!th]
\blue{
\footnotesize
  \centering
  \caption{Model prediction accuracy and classification metrics trained on spectrometer data from 28 Mar 2024}
  \label{tab:SVM_pred}
  \begin{minipage}{\columnwidth}
    \centering
    \begin{tabular}{|c|c|c|c|c|c|}
        \hline
        Date        & Accuracy & Precision & Recall & F1-Score & P-Value \\ \hline
        3 Apr 2024  & 79.4\%   & 100.0\%   & 58.9\% & 74.1\%   & 0.000   \\ \hline
        12 Apr 2024 & 75.6\%   & 68.5\%    & 94.4\% & 79.4\%   & 0.000   \\ \hline
        18 Apr 2024 & 77.8\%   & 98.1\%    & 56.7\% & 71.8\%   & 0.000   \\ \hline
    \end{tabular}%
    \captionsetup{justification=centering}
    \caption*{(a) Testing on two groups (control, salinity)}
  \end{minipage}
  \hspace{0.05\columnwidth}
  \begin{minipage}{\columnwidth}
    \centering
    \begin{tabular}{|c|c|c|c|c|c|}
    \hline
    Date        & Accuracy & Precision & Recall & F1-Score & P-Value \\ \hline
    3 Apr 2024  & 45.6\%   & 35.3\%    & 45.6\% & 38.2\%   & 0.000   \\ \hline
    12 Apr 2024 & 54.1\%   & 60.7\%    & 54.1\% & 53.3\%   & 0.000   \\ \hline
    18 Apr 2024 & 50.7\%   & 65.6\%    & 50.7\% & 46.5\%   & 0.000   \\ \hline
    \end{tabular}%
    \captionsetup{justification=centering}
    \caption*{(b) Testing on three groups (control, salinity, PRR)}
  \end{minipage}
  }
\end{table}


Second, we tested our model's predictive capability by training it on data from an earlier date and using it to predict treatment groups on later dates. As shown in Table~\ref{tab:SVM_pred}, this experiment was conducted under two scenarios: (a) distinguishing between control and salinity groups, and (b) classifying among control, salinity, and PRR groups. For the two-group classification, our model achieved up to 79\% accuracy with high statistical significance ($p < 0.05$), demonstrating reliable early detection capability. When extended to three-group classification, the accuracy decreased to 45-54\% but still maintained statistical significance. Note that while the accuracy of three-group classification might be deemed relatively low, this result is a significant improvement over other existing methods as none of them could provide meaningful results.

\blue{To better interpret the moderate three-class accuracy, we further analyzed class-wise precision, recall, and F1-score across the same testing periods. As shown in Table~\ref{tab:SVM_pred}, precision steadily improved from 35.3\% on April 3rd to 65.6\% on April 18th, indicating the model increasingly made confident and selective predictions. However, recall remained moderate (45.6\% to 54.1\%), suggesting that some true cases were undetected. The F1-score followed this pattern, rising from 38.2\% to 46.5\%. This reflects the difficulty in distinguishing between Salinity and PRR groups based on leaf spectral patterns alone. 

These results highlight that the decline in overall accuracy is not due to random behavior but rather reflects the underlying challenge of differentiating between the Salinity and PRR treatment groups based solely on leaf spectral measurements. Both stress conditions can produce overlapping spectral features, which increases classification difficulty. The higher precision but lower recall for later dates (e.g., April 18th) suggests that the model increasingly favors only highly confident predictions, potentially at the expense of recall. 
}



\section{Soil Sensor Data Analysis and Classification}\label{sec:soil_analysis}

This section analyzes the time-series soil conductivity and moisture data collected by low-cost soil sensors. We first assess existing classifiers, including traditional machine learning models, multivariate time-series networks, and recent state-of-the-art methods. Their limited accuracy on our dataset motivates us to develop a two-level hierarchical classifier. We then demonstrate the classification performance of our approach and evaluate the feasibility of deploying our approach to resource-constrained edge devices.

\subsection{Existing Methods}\label{sec:soil_baseline_methods}


For the analysis of our time-series soil sensor data, we evaluated several existing classification methods discussed in Sec.~\ref{sec:related_work_analysis}. Traditional machine learning methods (Random Forest, KNN, and SVM) were tested along with specialized time-series techniques (RNN with LSTM cells, ResNet, and MUSE~\cite{schafer2017multivariate}). To test state-of-the-art performance, we also implemented HIVE-COTE 2.0~\cite{hive}, MR-HYDRA~\cite{hydra}, and QUANT~\cite{quant}.

\begin{table*}[!th]
\caption{Comparison of our approach against existing generalized methods for Thomas and PP40 rootstocks}
\label{tab:classification}
\resizebox{\textwidth}{!}{%
\begin{tabular}{@{}cccccccccc@{}}
\toprule
\multirow{2}{*}{Category}                                                                & \multirow{2}{*}{Model}                                                              & \multicolumn{4}{c}{Thomas}                                                & \multicolumn{4}{c}{PP40}                                                  \\ \cmidrule(l){3-10} 
                                                                                         &                                                                                     & Accuracy         & Precision        & Recall           & F1-Score         & Accuracy         & Precision        & Recall           & F1-Score         \\ \midrule
\multirow{3}{*}{\begin{tabular}[c]{@{}c@{}}Traditional Machine \\ Learning\end{tabular}} & Random Forest                                                                       & 60.54\%          & 60.39\%          & 60.55\%          & 59.96\%          & 43.18\%          & 43.34\%          & 43.18\%          & 42.23\%          \\ \cmidrule(l){2-10} 
                                                                                         & KNN                                                                & 45.57\%          & 44.59\%          & 45.57\%          & 44.65\%          & 36.14\%          & 36.08\%          & 36.14\%          & 35.27\%          \\ \cmidrule(l){2-10} 
                                                                                         & SVM                                                                                 & 44.09\%          & 46.49\%          & 44.09\%          & 42.31\%          & 39.54\%          & 39.38\%          & 39.55\%          & 36.95\%          \\ \midrule
\begin{tabular}[c]{@{}c@{}}Multivariate Time \\ Series Classification\end{tabular}       & MUSE                                                                                & 60.13\%          & 61.30\%          & 60.12\%          & 59.82\%          & 42.50\%          & 42.72\%          & 42.50\%          & 41.93\%          \\ \midrule
\multirow{2}{*}{Neural Networks}                                                         & RNN (LSTM)                                                                          & 49.37\%          & 51.91\%          & 49.37\%          & 47.69\%          & 37.05\%          & 37.72\%          & 37.05\%          & 45.47\%          \\ \cmidrule(l){2-10} 
                                                                                         & ResNet                                                                              & 59.70\%          & 60.35\%          & 59.70\%          & 59.66\%          & 46.36\%          & 46.83\%          & 46.83\%          & 46.02\%          \\ \midrule
\multirow{3}{*}{State-of-the-Art}                                                        & HIVE-COTE 2.0                                                                       & 38.61\%          & 46.11\%          & 38.61\%          & 35.04\%          & 37.72\%          & 42.92\%          & 37.72\%          & 33.56\%          \\ \cmidrule(l){2-10} 
                                                                                         & MR-HYDRA                                                                            & 44.51\%          & 52.66\%          & 44.51\%          & 43.25\%          & 39.10\%          & 44.28\%          & 39.10\%          & 35.59\%          \\ \cmidrule(l){2-10} 
                                                                                         & QUANT                                                                               & 61.18\%          & 62.86\%          & 61.18\%          & 60.27\%          & 43.64\%          & 46.62\%          & 43.64\%          & 42.58\%          \\ \midrule
\textbf{Our Approach}                                                                    & \textbf{\begin{tabular}[c]{@{}c@{}}2-Level Hierarchical \\ Classifier\end{tabular}} & \textbf{86.50\%} & \textbf{87.49\%} & \textbf{86.50\%} & \textbf{86.73\%} & \textbf{74.32\%} & \textbf{75.01\%} & \textbf{74.32\%} & \textbf{73.98\%} \\ \bottomrule
\end{tabular}
}
\end{table*}

To evaluate these methods with our soil sensor data, we trained various models using the combined datasets from the first two periods (December 2023 to mid-March 2024) when stress symptoms were developing. Testing was performed on the third period data (mid-March to late April 2024) when symptoms were apparent.

Each model was evaluated using four standard classification metrics: accuracy, precision, recall, and F1-score. These metrics provide a more comprehensive understanding of the models' performance.
\begin{equation}
\text{Accuracy} = \frac{TP + TN}{TP + TN + FP + FN}
\label{eq:accuracy}
\end{equation}

\begin{equation}
\text{Precision} = \frac{TP}{TP + FP}
\label{eq:precision}
\end{equation}

\begin{equation}
\text{Recall} = \frac{TP}{TP + FN}
\label{eq:recall}
\end{equation}

\begin{equation}
\text{F1-Score} = 2 \cdot \frac{\text{Precision} \cdot \text{Recall}}{\text{Precision} + \text{Recall}}
\label{eq:f1score}
\end{equation}


Table~\ref{tab:classification} reports the results. While existing methods achieved statistical significance in distinguishing between different treatment groups ($p \approx 0$), their performance remained somewhat limited. As shown in the table, the accuracies of existing methods were only 38.61-61.18\% accuracy for Thomas and 31.2-46.36\% for PP40. The low F1-scores in particular indicate limited reliability in identifying stressed plants across both rootstocks. This highlights the need for a more tailored approach.

\begin{figure}[!th]
  \centering
  \includegraphics[width=\linewidth]{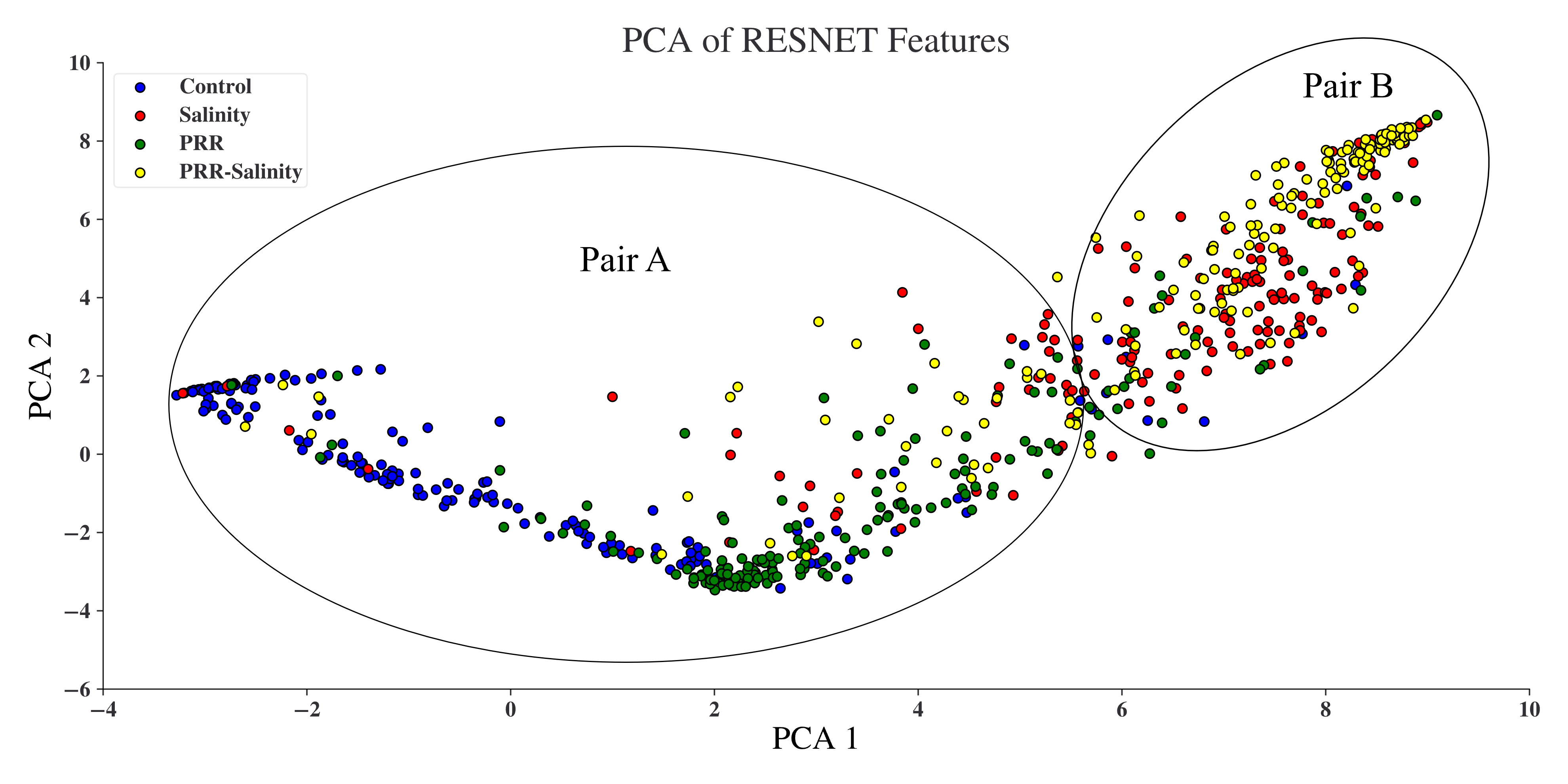}
  \caption{PCA of the features of a trained ResNet model}
  \label{fig:pca}
\end{figure}

\begin{figure}[!th]
  \centering
  \includegraphics[width=.8\linewidth]{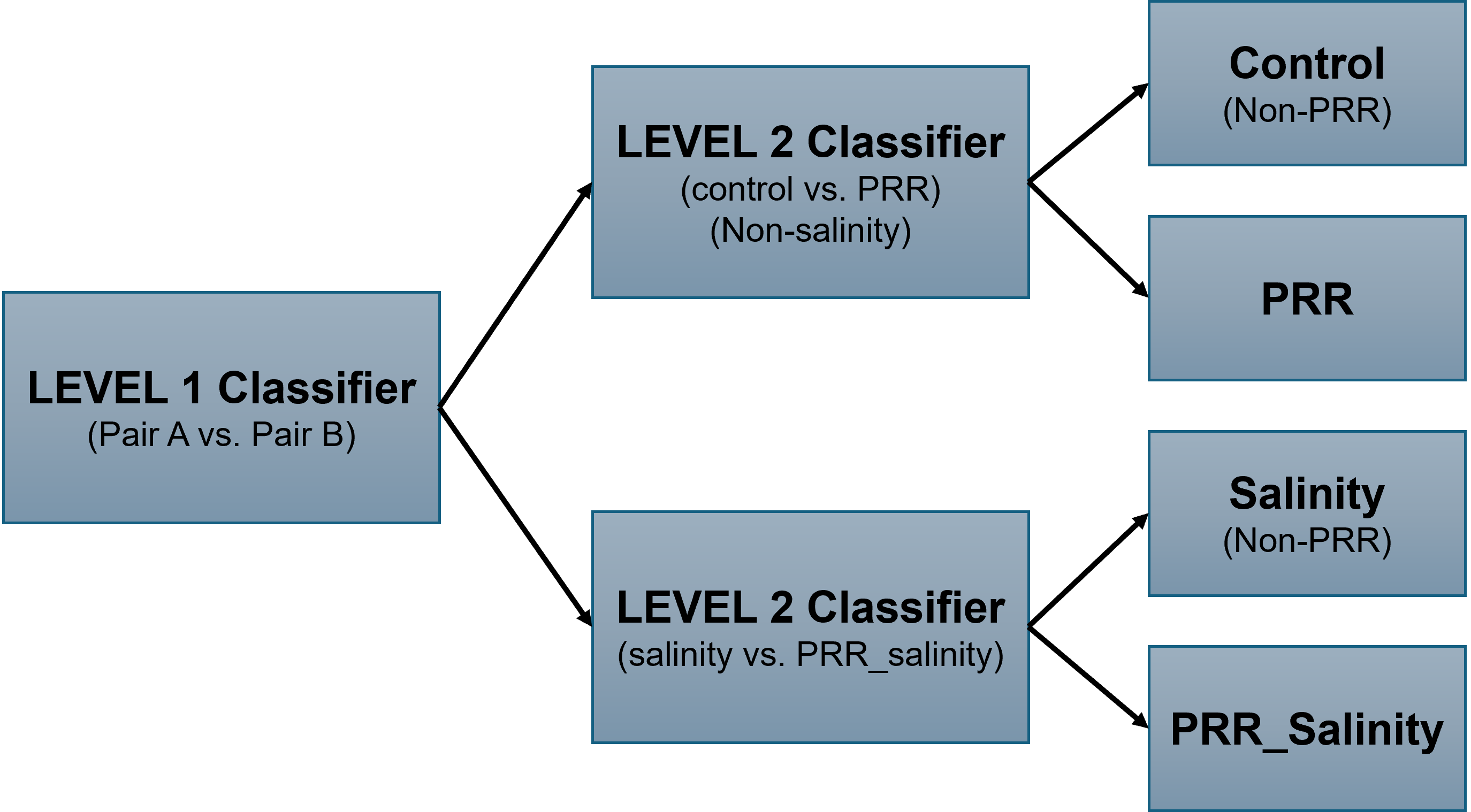}
  \caption{Proposed two-level hierarchical classifier}
  \label{2level}
\end{figure}

\subsection{Our Approach: Two-level Hierarchical Classifier}

To understand the limitations of existing methods, we analyzed the features extracted from the last layer of a trained ResNet model using Principal Component Analysis (PCA). As shown in Fig.~\ref{fig:pca}, the features of certain treatment group pairs significantly overlap, particularly between salinity and combined salinity-PRR groups. This overlap explains the degraded performance of existing classification approaches which attempt to differentiate all classes simultaneously.

The PCA visualization revealed that despite the overlap between certain treatment groups, there exists a clear separation between two pairs of groups: Pair A (control and PRR), and Pair B (salinity and salinity-PRR). This natural grouping in the feature space motivated us to design a two-level hierarchical classifier that first distinguishes between these pairs, then classifies the specific treatment within each pair.

\blue{To quantitatively validate this grouping strategy, we computed clustering quality metrics on the PCA-reduced features. The Silhouette Score~\cite{ROUSSEEUW198753} was 0.341 and the Calinski-Harabasz Index (CHI)~\cite{Caliński01011974} was 668.9, both indicating moderate but meaningful separation between the two groupings. These results reinforce the visual separation observed in the PCA plot (Fig.~\ref{fig:pca}) and support the effectiveness of our hierarchical design.}

Fig.~\ref{2level} illustrates our two-level classification approach. The first level (salinity-based classification) distinguishes between salinity-affected and non-salinity-affected plants, while the second level (PRR-based classification) further differentiates plants within each group based on PRR presence. For the level-1 classifier, we chose ResNet as it showed the best performance among baseline methods in distinguishing between these two categories. For level-2 classification within each pair, we explored different combinations of features and traditional machine-learning methods, i.e., Random Forest, K-Nearest Neighbors (KNN), and SVM, due to their low computational cost and effectiveness in binary classification problems, and ultimately selected Random Forest.

To build a reliable level-2 classifier, we extracted meaningful features from our time-series data that could effectively capture both temporal patterns and statistical properties. Using them, we constructed and tested two different feature sets to understand which ones play as key features for soil sensors.

Our first feature set includes four metrics: mean, standard deviation, skewness, and kurtosis. This combination captures a comprehensive profile of central tendency (mean), variability (standard deviation), and distributional shape (skewness and kurtosis), which has the potential to identify physiological responses to different stressors. On the other hand, our second feature set includes only two metrics: skewness and kurtosis. The reason behind this choice is to evaluate whether a simpler model (only higher-order statistical moments without basic descriptions) can also distinguish stressors well.

For both feature sets, we computed metrics for each daily window, enabling a time-sensitive analysis of plant responses as stress conditions developed. We then trained Random Forest, KNN, and SVM models with these feature sets and tested their performance on both pairs of treatment groups: (Control and PRR) and (Salinity and Salinity-PRR).

\subsection{Experimental Results}\label{sec:soil_analysis_results}

Following the same process as in the evaluation of baseline methods in Sec.~\ref{sec:soil_baseline_methods}, we trained our two-level hierarchical classifier using the combined datasets from the first two periods (December 2023 to mid-March 2024) and tested it on the third period (mid-March to late April 2024). As shown in Table~\ref{tab:classification}, our approach achieved significantly higher accuracy than baseline methods, reaching 86.5\% accuracy and an F1-score of 86.73\% for Thomas, and 74.32\% accuracy with a 73.98\% F1-score for PP40. 
For comparison, the state-of-the-art models such as HIVE-COTE 2.0, MR-HYDRA, and QUANT only achieved  38.61\%, 44.51\%, and 61.18\% accuracies, with corresponding F1-scores of 35.04\%, 43.25\%, and 60.27\% for Thomas, and 37.72\%, 39.10\%, and 43.64\% accuracies, with F1-scores reached only 33.56\%, 35.59\%, and 42.58\% for PP40, respectively.




Since the Thomas rootstock is susceptible to both salinity and PRR, our classifier was able to accurately distinguish between \blue{these stressors, as the corresponding soil moisture and conductivity signals reflected clear physiological changes.} However, the accuracy dropped for PP40, as this rootstock is tolerant to salinity and resistant to PRR. \blue{This tolerance delays the detection of physiological symptoms in the soil data during the early and middle stages of treatment, resulting in subtler changes in soil moisture and conductivity patterns. As a result, the separability between control and stressed treatments is reduced, leading to the observed decrease in classification performance.} Despite this variation in accuracy, we verified that our models achieved statistical significance in all cases ($p \approx 0$).


\begin{figure}[!th]
  \centering
  \subfloat[Thomas]{%
    \includegraphics[width=\columnwidth]{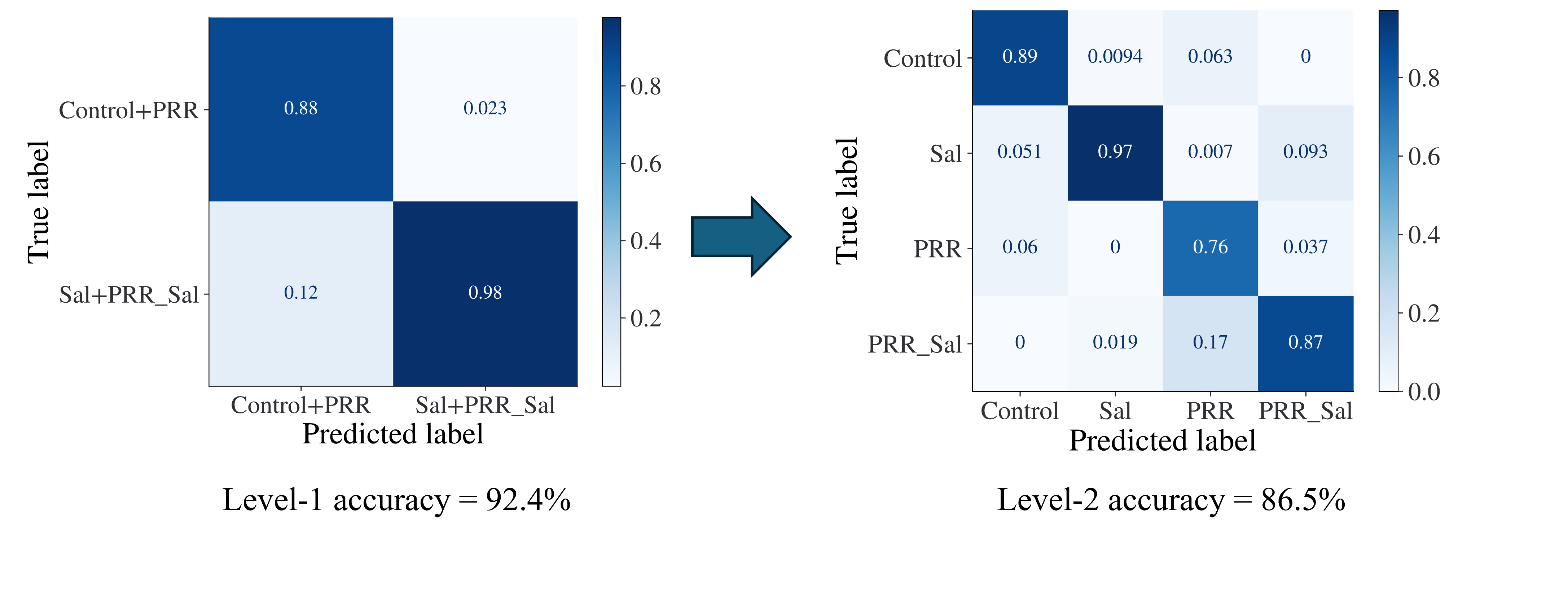} 
    \label{fig:thomas} 
  }

  \subfloat[PP40]{%
    \includegraphics[width=\columnwidth]{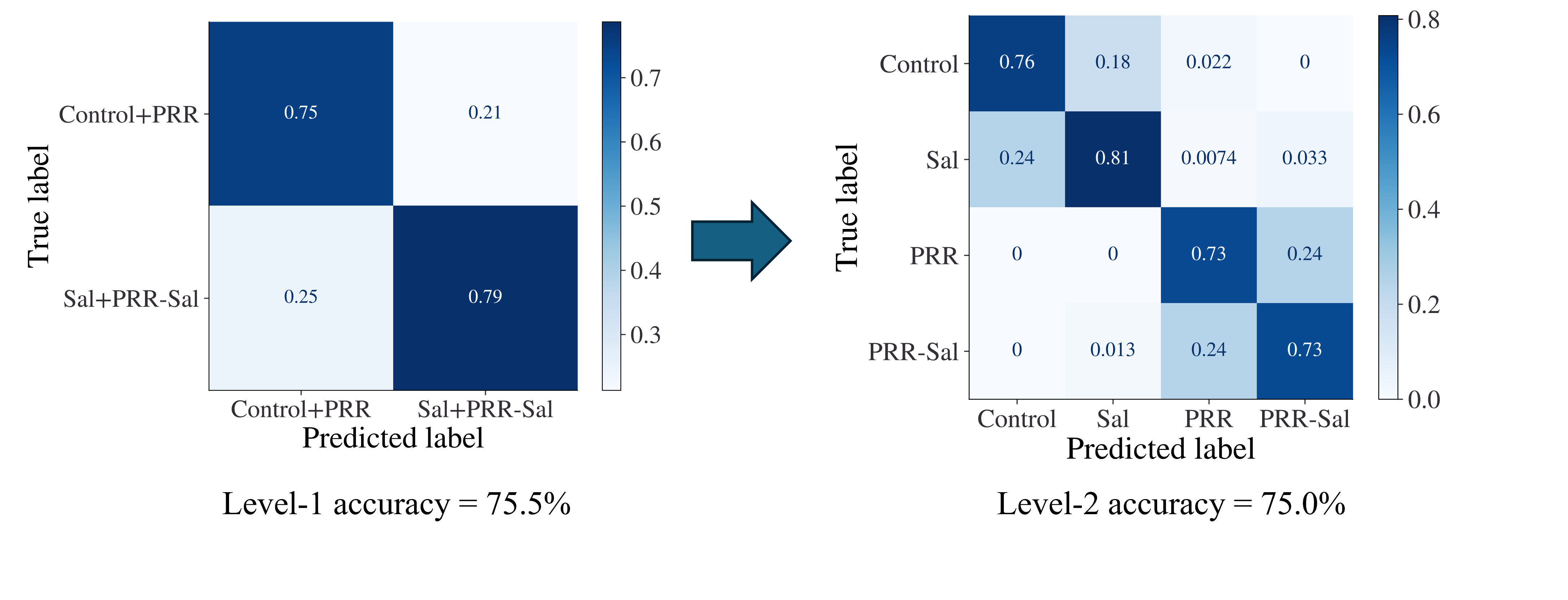} 
    \label{fig:pp40} 
  }

  \caption{Confusion matrices for Level-1 and 2 of our proposed classifier for Thomas and PP40 rootstocks}
  \label{fig:cm}
\end{figure}

The confusion matrices in Fig.~\ref{fig:cm} provide detailed insight into the performance at each level of our classifier approach. For level-1, the classifier effectively distinguished between Pair A (Control and PRR) and Pair B (Salinity and Salinity-PRR) with high accuracy for both rootstocks, though performance was better for Thomas compared to PP40.
This is expected, as PP40 is known to be resistant to PRR and tolerant to salinity. \blue{These traits delay the physiological impacts that are detectable by soil sensors, making the stress patterns less distinct during the early classification periods. This contributes to the greater overlap observed between treatment groups and explains the reduced classification performance.}
When moving to level-2 (Random Forest was used for Fig.~\ref{fig:cm}), accuracy decreased as anticipated due to the accumulated error propagated from level-1.



For the level-2 classifier, we also evaluated the effect of different feature sets for each candidate model. As shown in Table~\ref{tab:features}, Random Forest achieved the best performance with just two features (our second feature set). Although the difference is not high, we selected Random Forest with skewness and kurtosis as the winner for our level-2 model due to its high accuracy and moderate computational resources compared to the other two models. Sec.~\ref{sec:discussion} will provide a detailed discussion on the performance differences between the two feature sets.


\begin{table}[!th]
\caption{Comparing two sets of features for three ML methods for the level-2 classifier for each pair of stressors}
\label{tab:features}
\centering
\resizebox{\columnwidth}{!}{%
\begin{tabular}{ccccc}
\hline
Pair                                       & \# of Features & Random Forest   & KNN & SVM             \\ \hline
\multirow{2}{*}{Control and PRR}           & 4                  & 97.8\%          & 98.2\%                 & 98.7\%\\
                                           & 2                  & \textbf{99.5\%} & \textbf{99.5\%}        & 95.2\%          \\ \hline
\multirow{2}{*}{Salinity and PRR-Salinity} & 4                  & 90.5\% & 90.1\%                 & 90.5\% \\  
                                           & 2                  & \textbf{94.6\%} & 93.6\%                 & 92.3\%          \\ \hline
\end{tabular}%
}
\end{table}

\begin{figure*}[t]
    \centering
    \begin{subfigure}[t]{0.45\textwidth}
        \centering
        \includegraphics[width=\linewidth]{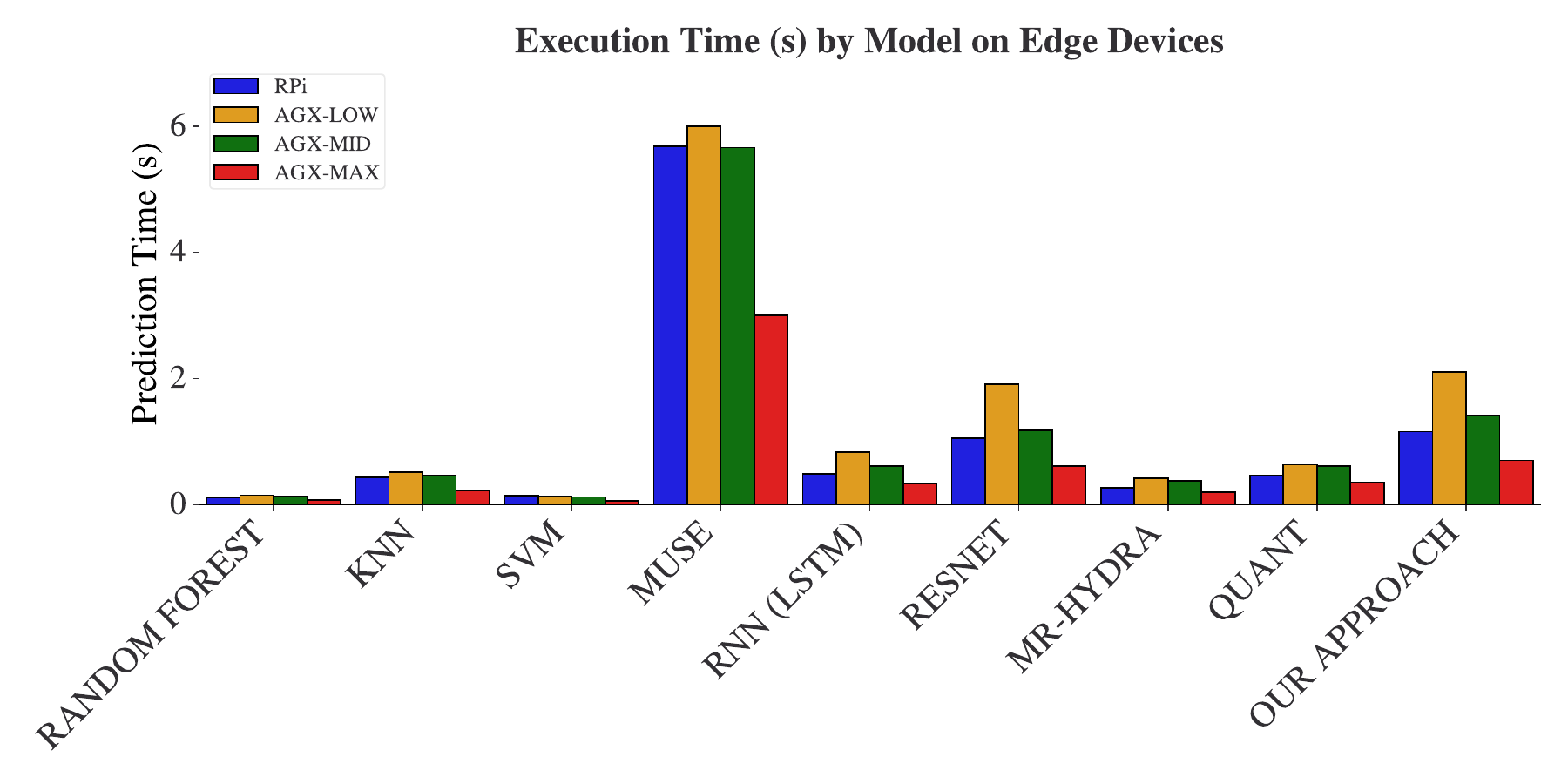}
        \caption{Execution Time}
        \label{fig:rpi_execution}
    \end{subfigure}
    \hfill
    \begin{subfigure}[t]{0.45\textwidth}
        \centering
        \includegraphics[width=\linewidth]{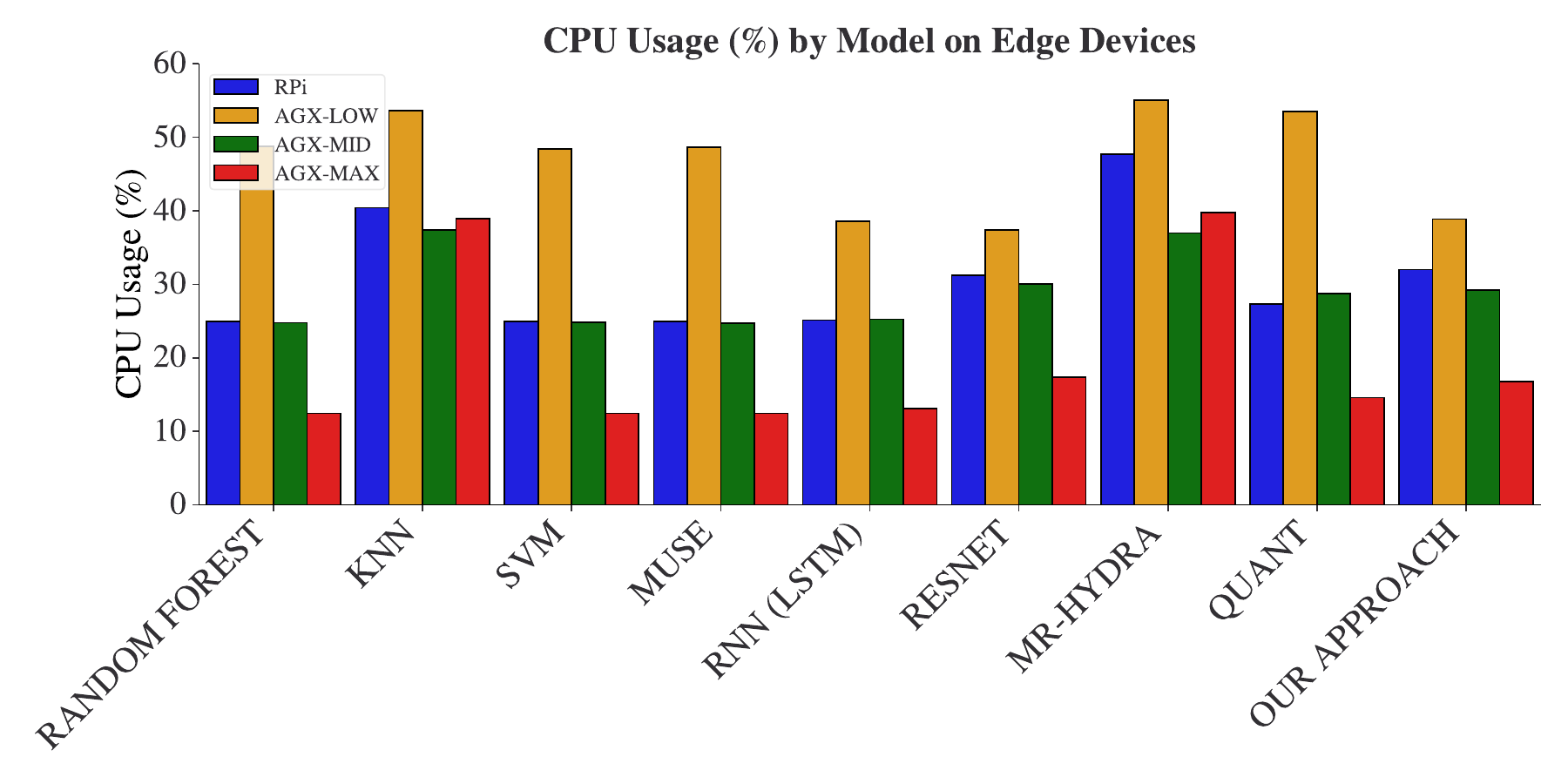}
        \caption{CPU Usage}
        \label{fig:rpi_cpu}
    \end{subfigure}

    \vspace{0.5em}  

    \begin{subfigure}[t]{0.45\textwidth}
        \centering
        \includegraphics[width=\linewidth]{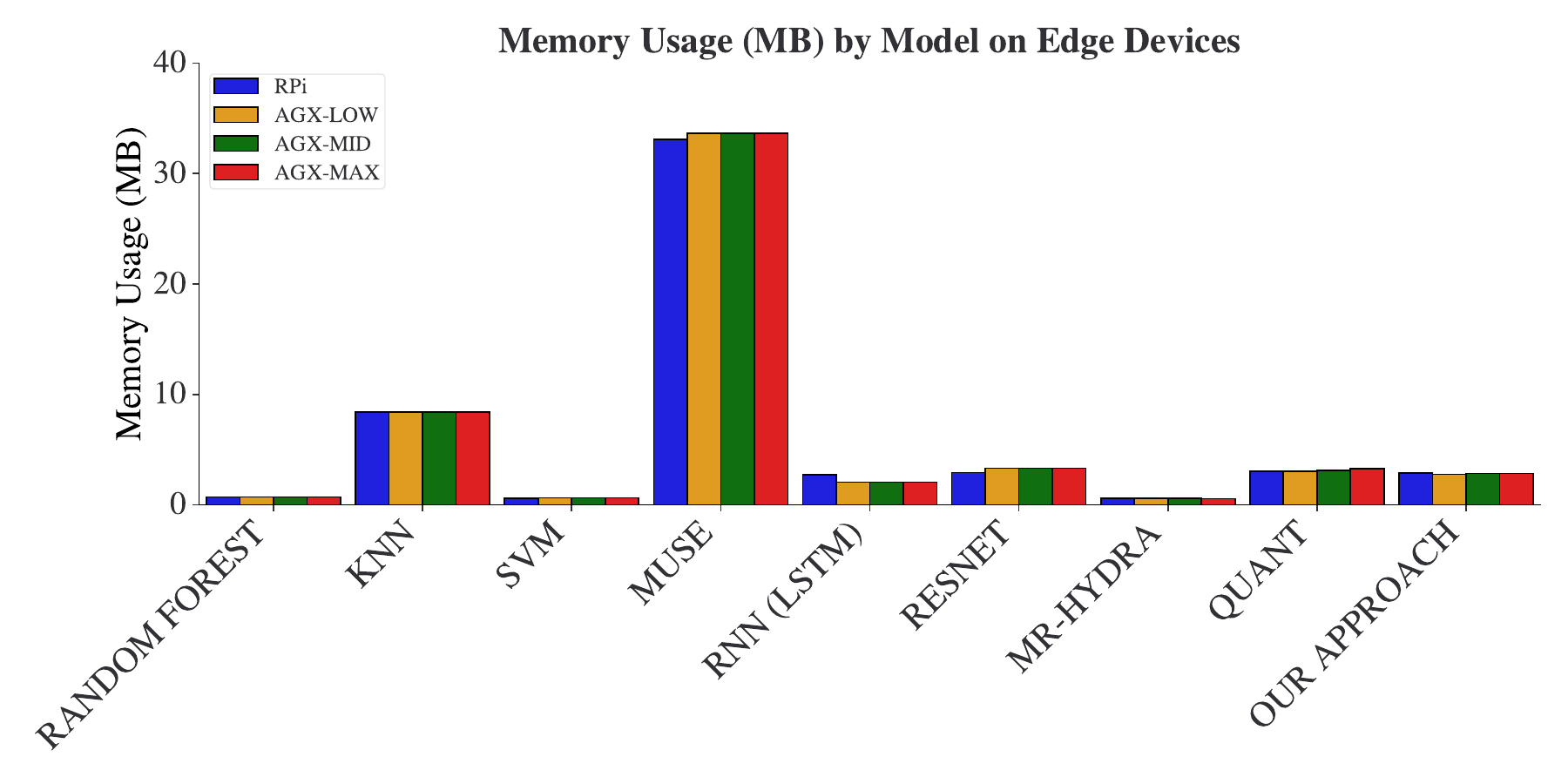}
        \caption{Memory Usage}
        \label{fig:rpi_memory}
    \end{subfigure}
    \hfill
    \begin{subfigure}[t]{0.45\textwidth}
        \centering
        \includegraphics[width=\linewidth]{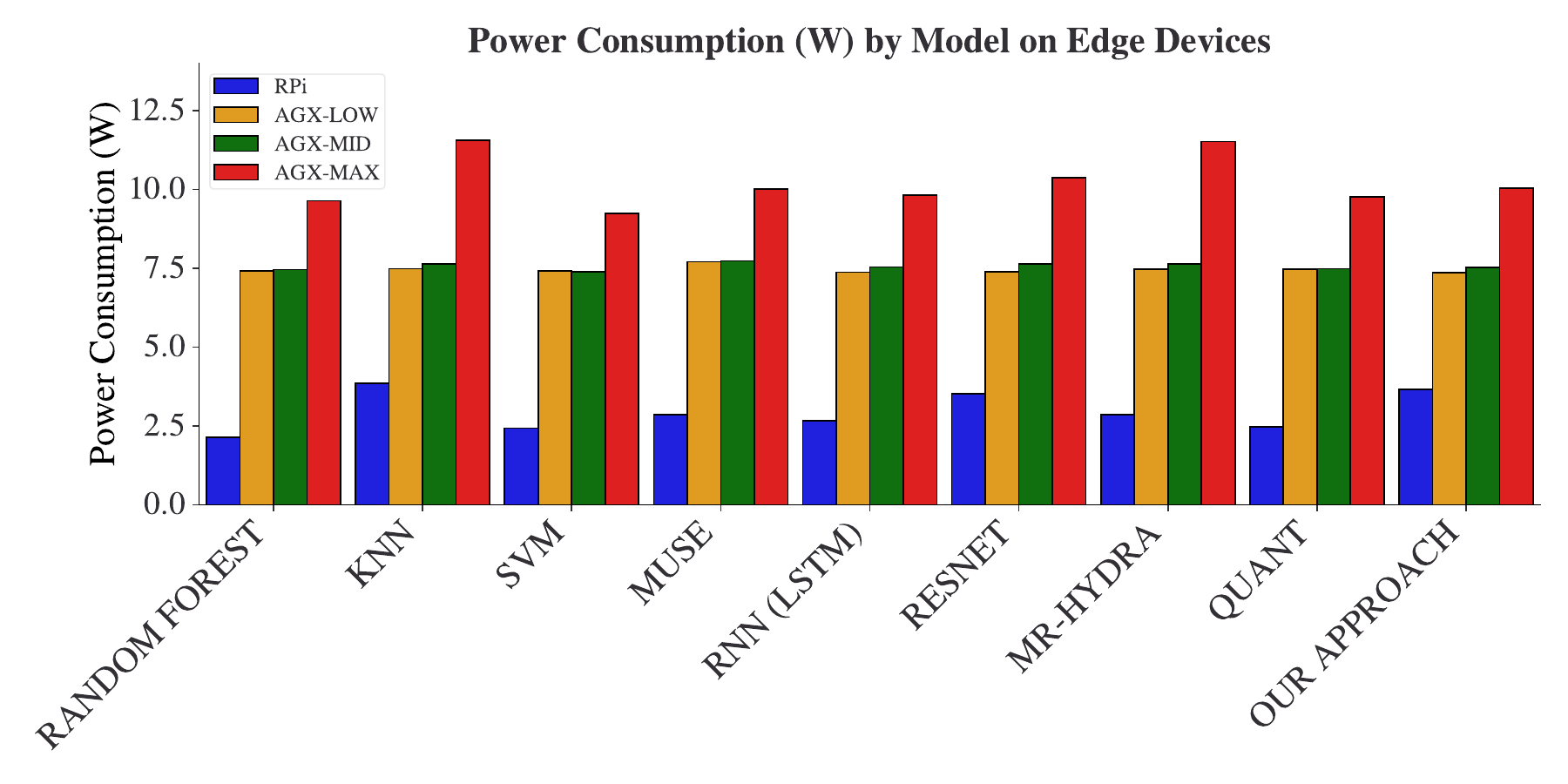}
        \caption{Power Consumption}
        \label{fig:rpi_power}
    \end{subfigure}

    \caption{\blue{Performance comparison of classifiers on Raspberry Pi 4B and NVIDIA Jetson AGX Xavier using different metrics: execution time, CPU usage, memory usage, and power consumption.}}
    \label{fig:rpi_performance}
\end{figure*}

\begin{table}[!th]
\centering
\footnotesize
\caption{\blue{Classification accuracy of Control group across 3 testing periods}}
\label{tab:control_stability}
\resizebox{\columnwidth}{!}{%
\blue{
\begin{tabular}{|c|c|c|}
\hline
\textbf{Period} & \textbf{ResNet (Level-1)} & \textbf{Random Forest (Level-2)} \\ \hline
Late Feb - Mid Mar 2024 & 84\% & 70\% \\ \hline
Mid Mar - Early Apr 2024 & 87\% & 80\% \\ \hline
Early Apr - Late Apr 2024 & 89\% & 82\% \\ \hline
\end{tabular}
}}
\end{table}

\blue{To verify the stability of our soil sensors and evaluate the potential impact of sensor aging or calibration drift, we conducted additional experiments focused on the Control group’s classification performance over time. 
We trained our classifiers to distinguish Control samples from treatment samples using data collected from late December 2023 to late February 2024, and tested them on samples collected during three later periods: late February to mid March 2024, mid March to early April 2024, and early April to late April 2024.
We evaluated this using both the ResNet-based model (Level-1) and the Random Forest classifier (Level-2) of our hierarchical pipeline. The results are summarized in Table~\ref{tab:control_stability}.
These results show no evidence of performance degradation across the study period; in fact, there were slight improvements over time (84\% to 89\% for ResNet, 70\% to 82\% for Random Forest), which could be due to more pronounced treatment effects over time. 
Both models maintained stable performance when classifying Control group samples, suggesting that sensor aging or calibration drift did not impact the reliability of our soil measurements. 
}

\subsection{Edge Device Performance Analysis}


Since our monitoring system continuously collects soil sensor data using an embedded edge device, it is worth evaluating the feasibility of deploying our two-level hierarchical classification approach in such a resource-constrained environment. 
To conduct this evaluation, we performed performance measurements on \blue{two popular embedded platforms: Raspberry Pi 4B and NVIDIA Jetson AGX Xavier.}
This experiment assesses execution time, CPU usage, memory usage, and estimated power consumption for our proposed approach in comparison to those of the baseline methods in Sec.~\ref{sec:soil_baseline_methods}. 
These metrics provide insights into the practicality of real-time monitoring and analysis in resource-constrained edge devices. Due to its extremely high memory footprint, HIVE-COTE 2.0 could not be executed on the Raspberry Pi and was therefore excluded from the edge performance evaluation.


The performance evaluation was conducted on data collected during the third period (mid-March to late April 2024), which was the same period as what we used for classification experiments in the previous section.  The dataset consists of 474 time windows, each containing 144 sensor readings, resulting in a total dataset size of 68,256. The reported execution time, CPU usage, memory usage, and estimated power consumption metrics are based on running each model across all these samples.

\subsubsection{Experimental Setup}
The evaluation involved executing each classification model on \blue{each edge device} and measuring the following parameters:

\begin{itemize}
    \item \textbf{Execution Time:} Average time required for the model to process a single input (daily window).
    \item \textbf{CPU Usage:} Average CPU utilization in percentage during execution.
    \item \textbf{Memory Usage:} Peak memory usage of the model. 
    \item \textbf{Estimated Power Consumption:} Estimated power consumption of the model, using an external monitoring tool, while processing an input.
\end{itemize}

\blue{Raspberry Pi 4B (RPi) has 8 CPU cores, each running at up to 1.8 GHz by default. We used this default power setting for RPi during experiments. NVIDIA Jetson AGX Xavier (AGX) is a much powerful edge platform than RPi, offering three different power modes: (i) MAX mode with 8 CPU cores running at up to 2.26 GHz, (ii) MID mode with 4 cores at 1.19 GHz, and (iii) LOW mode with 2 cores at 1.19 GHz. We tested all four configurations (1 RPi and 3 AGX power settings), as they can show energy-performance trade-offs.}


\blue{To ensure consistent and reliable performance metrics across platforms, each model was executed 20 times per hardware configuration. The reported execution time, CPU usage, memory usage, and power consumption values represent the average over these 20 runs. This methodology minimizes measurement noise and accounts for runtime variability.}

\subsubsection{Results}

    
The results of the edge performance evaluation are presented in Fig.~\ref{fig:rpi_performance}. Overall, our approach maintains acceptable resource usage while significantly outperforming all other models in classification accuracy. 


\blue{On RPi, our approach demonstrates a competitive execution time (1.16s) while maintaining moderate CPU usage (32.0\%) and memory consumption (2.90MB). While it includes a ResNet-based Level 1 classifier, the overall performance remains comparable to the standalone ResNet (1.05 seconds, 2.91MB), but with significantly higher classification accuracy. In contrast, the LSTM model is faster at 0.5 seconds but offers lower accuracy relative to its resource consumption. Traditional machine learning models such as Random Forest (0.11s) and SVM (0.14s) are more lightweight, yet they fall short in classification performance (e.g., 60.5\% and 44.1\% vs. our 86.5\% for the Thomas rootstock, as shown in Table~\ref{tab:classification}), making them less ideal for practical deployment. Additionally, our approach maintains an acceptable power consumption of 3.65W, which is on par with or slightly more than the other methods despite our superior accuracy.}

\blue{On AGX, the execution time of all methods, including our approach, is significantly reduced in MAX power mode (AGX-MAX in the figure). This is mainly due to the higher clock speed of AGX-MAX than that of RPi (2.26 GHz vs. 1.8 GHz). Memory usage remains similar to that of RPi, since the same program code runs on both platforms. The CPU usage of AGX-LOW is shown much higher than on RPi because it uses only 2 CPU cores running at a slower clock frequency. As expected, the power consumption of AGX increases with higher performance modes: from 7.36W in AGX-LOW to 10.03W in AGX-MAX. The comparison between RPi and AGX highlights the trade-off between computational efficiency and power consumption: RPi can sufficiently handle small to medium-sized environments like ours; AGX may be a better choice for larger-scale deployment scenarios where faster classification and higher throughput are required.}




Lightweight models like KNN, Random Forest, SVM, and MR-Hydra exhibit \blue{higher or similar CPU usage to our approach} despite shorter execution times because they perform computationally intensive operations that fully utilize the CPU but finish quickly. In contrast, models such as MUSE, ResNet, and our approach have \blue{moderate average CPU usage} but longer execution times; this is potentially due to that they involve more memory-bound operations, which cause the CPU to wait for data transfers. From these results, we conclude that our two-level classification approach effectively balances classification accuracy and resource efficiency, making it a viable option for resource-constrained, edge-based agricultural monitoring systems.

\section{Discussion}
\label{sec:discussion}

Among all leaf sensor measurements, reflectance data from the spectrometer proved most reliable and accurate in distinguishing between treatment groups, showing statistically significant differences among groups. However, our analysis revealed several important limitations in low-cost sensing approaches for plant stress detection.

The leaf temperature sensor's effectiveness was significantly compromised by environmental conditions. Fluctuations in ambient moisture and humidity levels in the field environment introduced substantial noise into temperature readings, making them difficult to isolate from temperature changes caused by plant stress. Similarly, leaf conductivity measurements showed poor reliability due to: (i) the sensor's sensitivity to exact positioning on the leaf surface, and (ii) variations in leaf morphology. These issues made it challenging to obtain consistent readings even from the same leaf over time.

These findings suggest that low-cost leaf temperature and conductivity (EC) sensors may not be suitable for field deployments. We recommend prioritizing spectrometer-based leaf reflectance measurements, which showed consistent and reliable performance when combined with our classification approach.  

Our two-level hierarchical classification approach also revealed important insights about low-cost sensing systems. The good performance of our approach was driven by the understanding of domain knowledge (e.g., grouping into treatment pairs). However, its generalization to scenarios with more treatment groups requires careful consideration. The primary issue lies in managing overlapping classes; as we add more treatment types, the likelihood of feature overlap increases, potentially degrading classification performance. Identifying and managing these overlapping classes based on domain knowledge will be crucial for successful deployment in complex field conditions.

The improved performance of the level-2 classifier when using only skewness and kurtosis (Sec.~\ref{sec:soil_analysis_results}; Table~\ref{tab:features}) suggests that mean and standard deviation may introduce redundancy rather than contributing meaningful discriminatory information. Since skewness and kurtosis capture variations in data distribution more effectively, they provide sufficient separation between stress conditions without additional complexity. 

The slight performance drop when incorporating all four features could be attributed to the curse of dimensionality, where unnecessary features introduce noise and reduce generalization. Reducing the feature set minimizes the risk of overfitting and allows the classifier to learn clearer decision boundaries, leading to improved classification accuracy with lower computational overhead. Furthermore, this helps reduce the model size and computational requirements, which are important for resource-constrained edge devices.

\blue{To assess robustness under potential deployment conditions, we repeated our evaluation using the Thomas rootstock dataset and introduced synthetic noise and missing values into the testing data. Specifically, 20\% of the soil sensor readings were randomly masked to simulate missing data, and Gaussian noise (mean 0, standard deviation 0.05) was added to the features. The model achieved 74.1\% with missing data and 86.3\% with added noise, compared to 86.5\% accuracy under normal testing. These results indicate robustness to sensor noise but greater sensitivity to missing data, which is consistent with the expected behavior for time-series soil monitoring systems.}


\section{Conclusion}

This paper presents a low-cost, sensor-based system for early detection of salinity and Phytophthora root rot (PRR) stressors in avocado trees. By integrating soil moisture level and electrical conductivity (EC) data from low-cost soil sensors with leaf reflectance data from a handheld sensor, our approach effectively differentiates between healthy plants, those under salinity stress, and those affected by PRR. For soil sensing, the proposed two-level hierarchical classifier successfully handled class overlap issues and achieved 75-86\% accuracy across different avocado genotypes (over 20\% higher than existing methods). For leaf sensing, we found that spectral reflectance is the most reliable indicator of stress in in-field conditions, whereas temperature and conductivity, despite their prevalent usage in controlled settings, suffer from inconsistencies due to environmental interference. 
Overall, our approach offers a cost-effective solution for early stress detection in avocado trees, with the potential to contribute to more accessible precision agriculture tools for improved crop management and productivity. 

\blue{There are several interesting directions for future work. For soil sensing, more advanced methods could be explored to manage class overlap in more complex scenarios and to generalize the proposed approach to other high-value crops. For leaf spectral reflectance analysis, additional dimensionality reduction techniques, such as domain-specific feature selection or sparse learning methods, would be worth considering. More targeted outlier detection methods might be incorporated to improve preprocessing reliability without discarding meaningful stress-related signals. Furthermore, the extension of our work to identifying unknown stress conditions would be an important future direction. 
}

\sloppy
\bibliographystyle{IEEEtran}

\bibliography{sample-base.bib}

\end{document}